\documentclass[10pt,twoside]{hsqcd}
\usepackage{epsf,amsmath}
\usepackage{graphicx}

\def\nostrocostrutto#1\over#2{\mathrel{\mathop{\kern 0pt \rlap
  {\raise.2ex\hbox{$#1$}}}
  \lower.9ex\hbox{\kern-.190em $#2$}}}
\def\gsim{\nostrocostrutto > \over \sim}   %greater or around...

\def\Journal#1#2#3#4{{#1} {\bf #2} (#4) #3}
\def\CPC{{ Comp. Phys. Comm.} }
\def\EPJC{{ Eur. Phys. J.} C}  
\def\IJMPA{{ Int. J. Mod. Phys.} A}

\def\JETPL{{  JETP Lett. }}

\def\MPLA{{ Mod. Phys. Lett.} A}

\def\NPB{{ Nucl. Phys.} B}
\def\PLB{{ Phys. Lett.} B}

\def\PRD{{ Phys. Rev.} D}

\def\PRp{{ Phys. Rep.} } %***********Khoze defined it as \PR

\def\ZPC{{ Z. Phys.} C}
 
\begin{document}

\title{QCD EVOLUTION OF HADRON AND JET \\
MULTIPLICITY MOMENTS
\thanks{
Presented at workshop ``Hadron Structure and QCD'' (HSQCD2004)
St. Petersburg, Repino, Russia, 18-22 May 2004.}}

%This work is
%supported by the High-Energy Physics  Foundation}}

\author{WOLFGANG OCHS\\
Max-Planck-Institut f\"ur Physik, F\"ohringer Ring 6\\
D-80805 M\"unchen, Germany\\
E-mail: wwo@mppmu.mpg.de}

\maketitle

\begin{abstract}
We describe recent applications of the MLLA evolution equations
to the calculation of mean multiplicities in quark and gluon jets and
the higher moments of hadron and sub-jet multiplicity
 distributions in $e^+e^-$-annihilation
as function of c.m.s. energy $Q$ and resolution parameter $y_{cut}$.
The transition from jets to hadrons 
with increasing jet resolution is considered.
\end{abstract}

%\markright{}

%\renewcommand{\@evenhead}

\markboth{\large \sl H.S. QCD-Author  \hspace*{2cm} HSQCD 2004}
{\large \sl \hspace*{1cm} TEMPLATE FOR THE HSQCD'04 PROCEEDINGS}

\section{Introduction}
Multiparticle production in hard collision processes is described
within QCD by combining the perturbative approach to the parton 
cascade evolution with a non-perturbative treatment of the 
transition towards the hadronic final state. The perturbative phase
is essentially determined by the QCD scale parameter $\Lambda$
and, possibly, the quark masses. 
The gluon bremsstrahlung which dominates the partonic cascade process 
with its collinear and soft singularities leads to the characteristic 
jet structure of the multiparton final state. The formation of partonic 
jets can be quantitatively studied by constructing jets explicitly using an
algorithm which combines partons into jets at an externally 
given resolution scale (parameter $Q_c$). 

In the calculation of jet production 
phenomena one relates  hadronic jets with
partons at the same resolution scale neglecting in general 
the effects of hadronization. It is an interesting question down to which 
scale one can follow this scheme of identifying a parton and a hadron jet
which we will adress below.

In a particularly simple ansatz for hadronization one assumes
that observables for the multiparticle final state are proportional to the
corresponding quantities for partons at a characteristic resolution scale $Q_0$, 
an idea which has been proposed
originally for single inclusive energy spectra and has been called
``Local Parton Hadron Duality'' (LPHD \cite{lphd}). Subsequently,
this idea has been applied to a wider range of phenomena including inclusive
multiparticle correlations and even quasi-exclusive processes (for reviews, see
\cite{dkmt2,ko}). The agreement with data generally increases with the 
accuracy of the calculation. The question arises whether this kind of 
agreement can be derived from a general principle or should the agreement be 
considered as largely accidental. Whereas there is no generally accepted
answer it is worth while to check this correspondence for more complex 
observables and at the same time to increase the accuracy of the predictions
aiming at a reliable phenomenology.

Here we will consider in particular the moments of multiplicity distributions
of hadrons and jets. We investigate to what extent
the jet observables can be connected with the corresponding hadron 
observables in the limit
\begin{equation}
Q_c\to Q_0.    \label{q0limit}
\end{equation}
Such a connection would suggest a 
one-to-one correspondence between hadrons and partons.
Clearly, such a correspondence cannot exist in general for any exclusive limit 
but only after a certain averaging in the sense of a dual description. 

The mean multiplicities and multiplicity moments 
in quark and gluon jets as function of primary energy 
and (sub-) jet resolution has been calculated using the evolution equation
in the Modified Leading Logarithmic Approximation (MLLA) \cite{lo,bfo}.
This equation has been presented originally at the Leningrad Winterschool 1984 
\cite{lewi}, is explained in 
\cite{dkmt2}, and can be considered as an extension of the well known 
DGLAP evolution equation  
towards small particle energies
 taking into account 
soft gluon coherence as realized in a probabilistic way by angular ordering
\cite{ao}. Whereas there have been various levels of analytical approximations 
to the solution of the MLLA evolution equation  
with increasing number of subleading logarithmic terms 
we find that the numeric 
solution of this equation yields quantitative agreement with the variety of 
global observables discussed here.

\section{Moments of multiplicity distribution}
Be $P_n$ the distribution of particle (parton) multiplicity in a jet.
Then we consider the unnormalized and normalized factorial moments $f_q$
and $F_q$
\begin{equation}
f_q=\sum_{n=0}^\infty n(n-1)\ldots (n-q+1)P_n, 
   \quad  F_q=f_q/N^q, \quad  N\equiv f_1
\label{fmom}
\end{equation}
with mean multiplicity $N$.
Furthermore, one introduces the cumulant moments
$k_q$ and $K_q$ which are used to measure the genuine correlations without  
 uncorrelated background in a multiparticle sample
\begin{equation}
k_q=f_q-\sum_{i=1}^{q-1} {q-1 \choose i} k_{q-i} f_i, \qquad K_q=k_q/N^q,
\label{kmom}
\end{equation}
in particular $K_2=F_2-1,\ K_3=F_3-3F_2+2$; for a Poisson distribution
$K_1=1,\ K_q=0$ for $q>1$.
The moments can be conveniently computed with the help of 
the generating function
\begin{eqnarray}
Z(u)&=& \sum_{n=1}^\infty P_n u^n\\
f_q&=& \frac{\partial^n Z(u)}{\partial u}|_{u=1},\qquad
   k_q\ =\ \frac{\partial^n \ln Z(u)}{\partial u}|_{u=1} 
\label{genfunction}
\end{eqnarray}
Of special interest are
the ratios 
\begin{equation}
H_q=K_q/F_q
\end{equation}
which have been predicted to show
an oscillatory behaviour at high energies \cite{dreminosc}
with the first minimum near $q\approx5$ at the mass of the $Z$ boson.
Such a minimum has been observed indeed in
$e^+e^-$ annihilations at SLC \cite{slacosc} and  at LEP
\cite{L3osc,mangeol} but the magnitudes of moments are found much smaller
than originally expected in \cite{dreminosc}.

\section{Perturbative QCD predictions}

Predictions on the global quantities as above can be obtained from the 
MLLA evolution equation for the generation function $Z(Y_c,u)$
and the initial condition at threshold
which read in the simplified world of gluodynamics without quarks
\begin{eqnarray}
\frac{d}{dY_c} Z(Y_c,u)&=& \int_{z_c}^{1-z_c} dz
   \frac{\alpha_s(\tilde{k_T})}{2\pi}P_{gg}(z)\times \nonumber\\
     && \hspace{1.5cm} \{Z(Y_c + \ln z,u)Z(Y_c + \ln (1-z),u) - Z(Y_c,u)\}
\label{Zevol}\\
Z(0,u)&=& u.
\label{Zbound}
\end{eqnarray}
In the general case there are two coupled equations for $Z_g$ and $Z_q$.
In (\ref{Zevol}) the evolution variable is 
$Y_c=\ln \frac{E}{Q_c}$ in the jet energy $E$ and resolution parameter $Q_c$.
The parameter $Q_c$ limits the transverse momentum from below
$\tilde k_T = \min (z,1-z)E > Q_c$. This restriction yields a parton cascade
with a minimum $k_T$ separation and can be compared with the jet ensemble
constructed from hadrons using the
so-called $k_T$- or ``Durham''-algorithm. The initial condition (\ref{Zbound})
sets the multiplicity to $N=1$ at threshold $E=Q_c$ and $F_q=0$ for $q>1$.

Asymptotic solutions can be obtained in the Double Logarithmic Approximation
(DLA) which includes only the dominant contributions from the collinear and 
soft singularities, i.e. the splitting function 
$P_{gg}(z)\sim 1/z$ in (\ref{Zevol}); 
the next to leading single log 
terms are included in the MLLA. Up to this order the results are complete; 
further logarithmic contributions beyond NLLO can be calculated, but they are 
not complete and neglect in particular process dependent large angle emissions.
 Nevertheless they improve the results considerably as they take into account 
energy conservation with increasing accuracy. The full 
solution of Eq. (\ref{Zevol}), corresponding to the summation of all logarithmic orders, can 
be obtained numerically. Alternatively, one may calculate results of the QCD 
cascade from a Monte Carlo generator, we compare here especially with ARIADNE
\cite{ARIADNE}, which is based on similar construction principles as  
Eq. (\ref{Zevol}).  

\section{Mean particle multiplicity in quark and gluon jets}
The multiplicities $N_g,N_q$ in gluon and quark jets can be
obtained from the MLLA evolution equations. At high energies one can write
\begin{equation}
N_g(Y) \sim  \exp\left(\int^Y\gamma(y)dy\right) \label{nasy}
\end{equation}
where the anomalous dimension $\gamma$ has the expansion in
$\gamma_0=\sqrt{2C_A\alpha_s/\pi}$
\begin{equation}
\gamma=\gamma_0\ (1-a_1 \gamma_0 - a_2 \gamma_0^2 - a_3 \gamma_0^3 \ldots),
\label{gamma}
\end{equation}
likewise the ratio of gluon and quark jet multiplicity
\begin{equation}
r\equiv \frac{N_g}{N_q}=\frac{C_A}{C_F}(1-r_1\gamma_0 -
r_2\gamma_0^2-r_3 \gamma_0^3 \ldots).
\label{rgq}
\end{equation}
with the colour factors $C_A=3$ and $C_F=\frac{4}{3}$. The coefficients
$a_i$ and $r_i$ can be obtained from the evolution equations.

The rise of parton
multiplicity in a quark jet is given in MLLA by\\
$N\propto \exp{[c_1\sqrt{\ln(E/\Lambda)}+c_2\ln\ln (E/\Lambda)]}$
and this formula describes well the data in $e^+e^-$ annihilation 
at LEP1 and LEP2 (for reviews, see \cite{dg,ob}
and \cite{hamacher}). 

\begin{figure}[t!] %[!thb]
\vspace*{8.5cm}
\begin{center}
\includegraphics{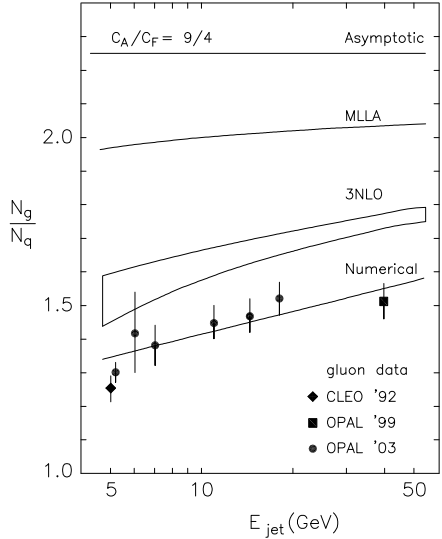}
%\centerline{\epsfxsize=2.9in\epsfbox{kim_mephi_lep.ps}}
\vspace{-1.0cm}
\caption[*]{ 
The ratio of the mean multiplicities in gluon jets and quark jets 
$N_g$ and $N_q$. Results from evolution equations of different order of 
approximation in comparison with experimental data obtained 
in $e^+e^-$-annihilation.
}
\end{center}
\label{fig:rgq}
\end{figure}

The role of higher logarithmic orders can be studied in the behaviour 
of the multiplicity ratio $r$ in (\ref{rgq}). The asymptotic limit $r=C_A/C_F$
acquires large finite energy corrections
 in  NLLO
\cite{ahm1,mw} and 2NLLO order \cite{gm,dreminosc}
\begin{eqnarray}
 r_1 &=& 2\left(h_1+\frac{N_f}{12N_C^3}\right) -\frac{3}{4}\\
 r_2 & =& \frac{r_1}{6} \left(\frac{25}{8}- \frac{3N_f}{4N_C} -
                               \frac{C_FN_f}{N_C^2}
           -\frac{7}{8}-h_2-\frac{C_F}{N_C}h_3+\frac{N_f}{12N_C} h_4\right)
\end{eqnarray}
with $h_1=\frac{11}{24},\ h_2=\frac{67- 6\pi^2}{36},\ 
h_3=\frac{4\pi^2-15}{24}$ and $ h_4=\frac{13}{3}$, also 3NLLO results have
been derived \cite{cdgnt}. Results from these approximations \cite{dg}
are shown in Fig.~1 %\ref{fig:rgq} together
together with the numerical solution of the MLLA evolution equations 
obtained in 1998 \cite{lo}
 which takes into account all higher order corrections from this equation
and fulfils the (non-perturbative) boundary condition (\ref{Zbound}).
All curves are absolute predictions, as the parameter $\Lambda$ (and $Q_0$ in
case of the numerical calculation)
is adjusted from the growth of the total particle multiplicity
in the $e^+e^-$ jets.
The slow convergence of this $\sqrt{\alpha_s}$ expansion can be seen
and there are still considerable effects beyond 3NLLO. The 
numerical solution is also in close agreement with 
the MC result at the parton level obtained \cite{opalmult} from the HERWIG MC 
above the jet energy 
$E_{\rm jet}>15$ GeV ($E_{\rm jet}=Q/2$ in $e^+e^-$ annihilation)
 and $\sim$ 20\% larger
at $E_{\rm jet}\sim 5$ GeV.
This overall agreement suggests that the 
effects not included in the MLLA evolution equation, 
such as large angle emission, are small.

These numerical results are also compared in Fig.~1 % \ref{fig:rgq} with 
with data from OPAL \cite{opalmult} where the data on gluon jets are derived from 
3-jet events in $e^+e^-$-annihilation. Note also that a proportionality constant
relating partons and hadrons
according to LPHD drops in the ratio $r$.
Recent results from DELPHI 
\cite{delphimult,hamacher} fall slightly below the curve by about 20\% at the lowest
 energies but converge for the higher ones; 
the CDF collaboration comparing quark and gluon jets at high $p_T$
in $pp$ collisions \cite{pronko} finds the ratio $r$ in the range 
$5<E_{\rm jet}<15$ GeV a bit larger, closer to the 3NLLO prediction,
 but with larger errors 
and therefore still consistent with the LEP results.

An alternative calculation is based on the colour dipole model which treats 
the evolution of dipoles 
in NLL approximation and includes recoil effects \cite{eden}. It provides 
a good description of the data but includes an extra (non-perturbative)
 parameter which allows to adjust a low energy point. Whether such
a non-perturbative input is definitely required by the DELPHI data depends 
in particular on the theoretical uncertainty
in the extraction of $N_g$ from 3-jet events.

\section{Multiplicity moments of hadrons and sub-jets}
We now differentiate between  sub-jets and hadrons in a jet
of hard scale $E=Q/2$ in $e^+e^-$ annihilation.
Sub-jets are defined by the resolution scale $Q_c$ ($k_T>Q_c>\Lambda$), hadrons
are related to partons at scale $Q_0$ ($k_T>Q_0$) and experience shows that
$Q_0\approx \Lambda$.

In the DLA the multiplicity of partons at resolution $Q_c$ in a jet
can be obtained analytically from (\ref{Zevol}) with (\ref{Zbound})
in terms of modified Bessel functions 
\begin{equation}
N_g (Y)=\beta \sqrt{Y+\lambda_c}
   \{ K_0(\beta \sqrt{\lambda_c}) I_1(\beta \sqrt{Y+\lambda_c}) +
      I_0(\beta \sqrt{\lambda_c}) K_1(\beta \sqrt{\lambda_c}) \}. 
\label{dlarun}       
\end{equation}
where $\beta^2=\frac{16 N_C}{b}$ with $ b=\frac{11}{3} N_C-\frac{2}{3}n_f$
and $\lambda_c=\ln\frac{Q_c}{\Lambda}$. At high energies $I_n,K_n$ rise 
exponentially and one obtains 
in two limits for resolution $Q_c$ 
\begin{eqnarray}
\textrm{at high resolution}&  (Q_c\to Q_0):  &    N\sim
(\beta^2Y)^{1/4} \ln\left(\frac{2}{\beta\sqrt{\lambda_c}}\right)
     \exp{\sqrt{\beta (Y+\lambda_c)}} \phantom{abc}  \label{hres}\\
\textrm{at low resolution } &  (Q_c\to E):  & 
    N\to 1,
\label{resolvedla}
\end{eqnarray}
where we also used $K_0(z)\simeq \ln(2/z)$ for small $z$. At high resolution
for $Q_c~\to~\Lambda$ ($\lambda_c\to 0$) the parton 
multiplicity diverges logarithmically because
of the Landau pole appearing in the running coupling. The pole is shielded
by the cut-off $Q_c=Q_0$ and at this value the parton multiplicity $N$
reaches the hadron multiplicity $N_h$ according to the LPHD prescription
(up to an overall constant $K$). 

\begin{figure}[t!]%[hbt]
\begin{center}
\hspace{-1cm}\includegraphics[angle=-90,width=12.5cm]{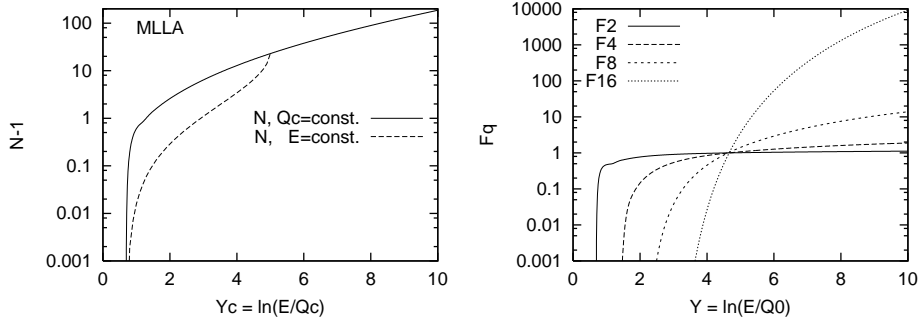}
\end{center}
\vspace{-4.5cm}
\caption[*]{
Parton multiplicity $N$ in MLLA in a single gluon jet 
 vs energy for fixed $Q_c=Q_0$ 
(representing hadrons, full line) and at fixed energy $Y_0=\ln(E/Q_0)=5$
(LEP-1 energy) for variable jet resolution $Q_c$ (l.h.s.)
and the
factorial moments $F_q$ vs. energy $Y$ (r.h.s.). Numerical solutions
of evolution eq., taken from \protect\cite{bfo}.
}
\label{hqmlla-nf}
\end{figure}     

The multiplicity in MLLA comes out considerably
 smaller than in DLA, however the dependences on $E$ and $Q_c$ are
 qualitatively 
the same. It is shown in Fig. \ref{hqmlla-nf}: for fixed $Q_c=Q_0$ the 
multiplicity $N$ starts with $N=1$ at threshold and rises
 $\sim \exp(c\sqrt{\ln E})$; at fixed energy $E$ we find $N=1$ for $Q_c=E$ 
according to (\ref{resolvedla}) whereas $N$ rises rapidly
for $Q_c\to \Lambda$  
and ultimately approaches the upper curve at $Q_c=Q_0\gsim \Lambda$. 
The splitting between the upper and lower curve is entirely due to the 
running of the coupling $\alpha_s$.

\begin{figure}[t!]%[hbt]
\begin{center}
\hspace{-2cm}\includegraphics[angle=-90,width=11cm]{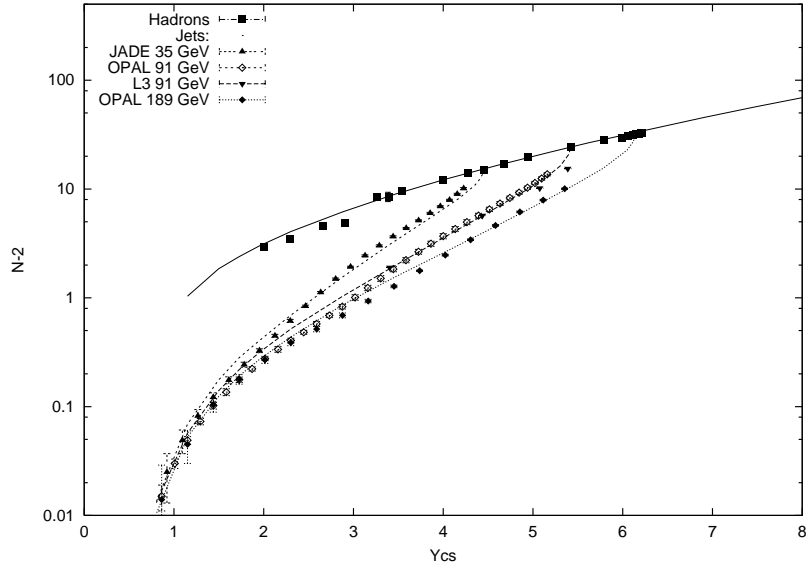}
\end{center}
\vspace{-0.3cm}
\caption[*]{
Multiplicity $N$ of hadrons
(taken as $N_{ch}\times1.25$) 
%using data \cite{slacmult,argus,jade,tasso2,hrs,amy,%
%alephgp,delphigp,l3gp,opalgp} and LEP averages from \cite{dg})
and multiplicity of jets %\cite{opalpfeifen,L3osc} 
in $e^+e^-$ annihilation at three energies $Q$
together with parton Monte Carlo results (parameters 
$\Lambda,Q_0$ fitted) as function of 
$Y_{cs}=\ln(Q^2/(Q_c^2+Q_0^2))$, Fig. from \protect\cite{bfo}.
}
\label{multiplicity}
\end{figure} 

The full numerical
solution of the MLLA evolution equations has been studied already in \cite{lo}.
Similar results are obtained from the ARIADNE MC 
at the parton level \cite{ARIADNE} with 
readjusted two parameters $\Lambda,\ Q_0$ according to the duality approach 
(``ARIADNE-D''). They are shown in Fig. \ref{multiplicity} 
in comparison with the experimental 
hadron multiplicities (upper curve) and the jet multiplicities 
in $e^+e^-$ annihilation represented as superposition of 2 single jets
($N=2$ at threshold). 
The normalization of the hadron data is adjusted to the calculation; 
taking into account that the data do not include neutrals
one finds that the proportionality factor $K$ between 
multiplicities of hadrons and partons at scale $Q_0$ is close to $K=1$
in agreement with previous findings \cite{lo}. In
the comparison between data and calculations we used the variable
$Y_{cs}$; it takes into account that in the experimental (and MC) 
jet algorithm
the hadrons are resolved at $Q_c=0$ but in the analytical MLLA calculation 
at $Q_c=Q_0$. 

A satisfactory overall description of the data can be obtained, especially
in the transition region between jets and hadrons, with parameters
$\Lambda=400$ MeV and $Q_0=404$ MeV.
Some disagreement with data of the jet curves
occurs around $Q_c>10$ GeV which may
be related to $b\bar b$ production, not included in the calculation.
The differences between the various curves are 
determined by the behaviour of the running coupling: for fixed coupling 
all curves would coincide and behave like a power $N\propto (Q/Q_c)^\alpha$.
The rapid variation of the jet curves at their upper end comes 
from the closeness of the parameters $Q_0$ and $\Lambda$.

\begin{figure}[t!]%[hbt]
\begin{center}
%\begin{minipage}[b]
%\centering
\hspace{-2cm}\includegraphics[angle=-90,width=8cm]{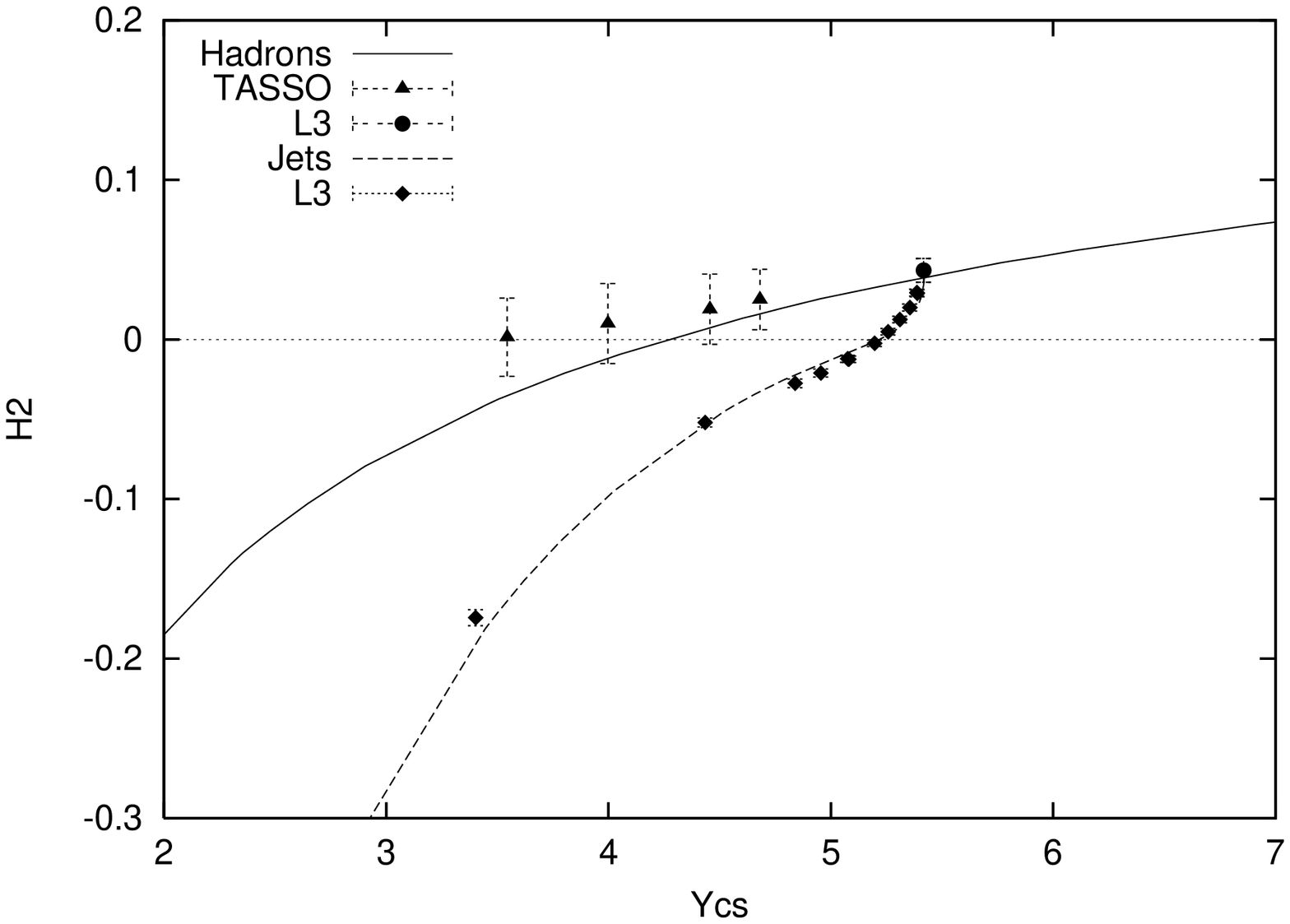}\\%[0.5cm] %\hspace{1cm}%
%\end{minipage}[b]% 
%\begin{minipage}[b]
%\centering
\hspace{-2cm}\includegraphics[angle=-90,width=8cm]{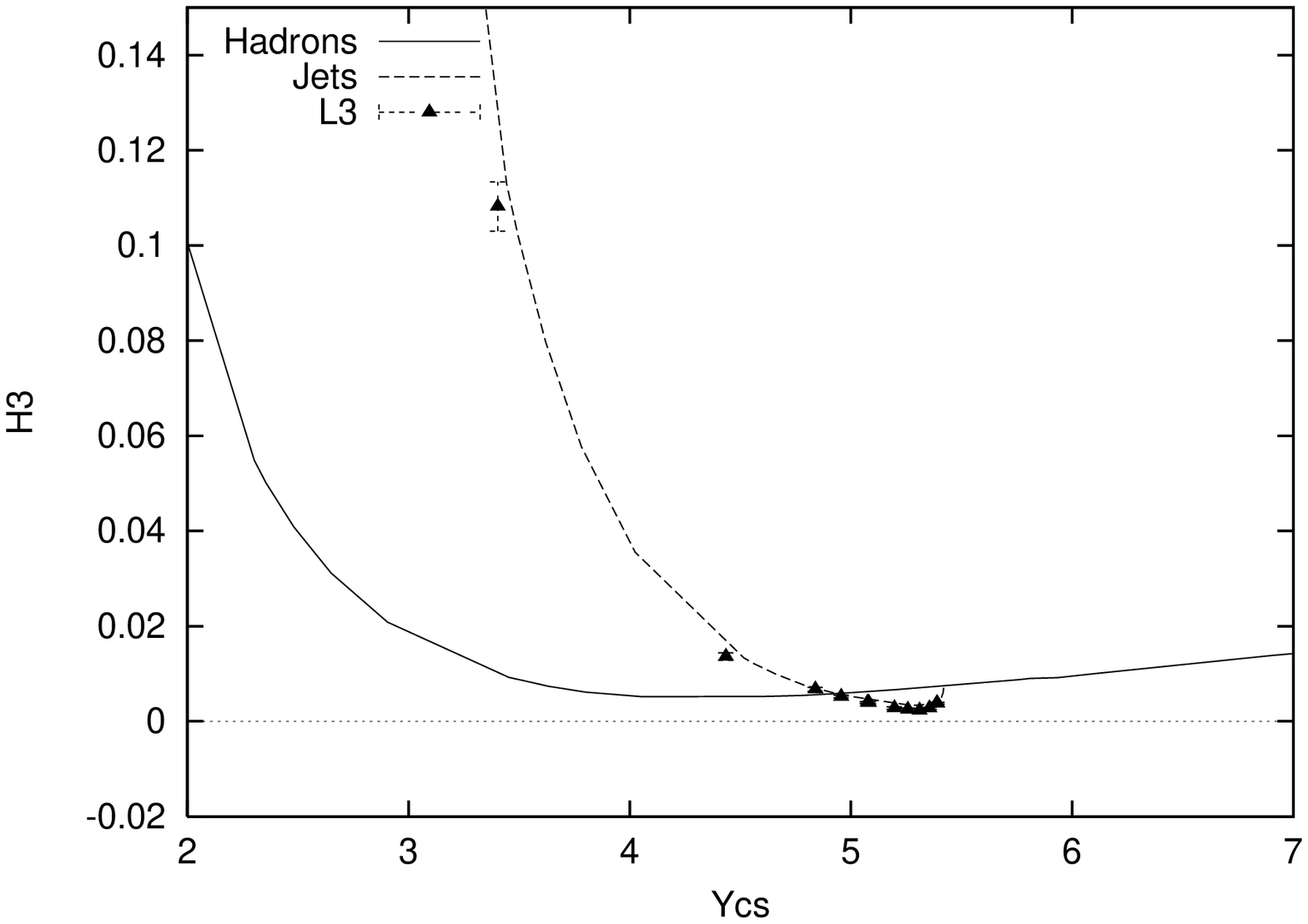}\\% 10cm  
%\end{minipage}[b]
\end{center}   
\vspace{-0.3cm}
\caption[*]{    
Ratio of moments $H_q$ for hadrons as function of energy $Q=2E$ (fixed $Q_0$)
 and for jets as function of resolution $Q_c$ at $Q=91$ GeV using 
the variable $Y_{cs}$ as in Fig. \ref{multiplicity} 
in comparison with ARIADNE-D MC, from \protect\cite{bfo}.}
\label{hq2-3exp}
\end{figure}

Next we turn to the higher multiplicity moments. The energy dependence 
of the factorial moments $F_q$ in MLLA is also shown in Fig. \ref{hqmlla-nf}.
The calculation takes into account energy conservation, therefore the 
threshold for production of $q$ particles is shifted towards 
$E_{\rm thr}=qQ_0$. Remarkably, all $F_q$ curves cross to good approximation 
at $F_q=1$, a Poissonian point. For a Poissonian, the kumulant moments $K_q$
and therefore the ratios $H_q=K_q/F_q$ vanish for $q>1$. 
At threshold one finds the rapid oscillations with $q$
of the kumulants $K_q=(-1)^{q-1}(q-1)!$, they continue up to the Poissonian 
point with decreasing amplitudes. At higher energies, the oscillation length
of $K_q,H_q$ increases  and should approach asymptotically
the DLA limit $H_q\sim 1/q^2$  with all kumulants positive.

In Fig. \ref{hq2-3exp} we show as representative 
examples the moment ratios $H_2$ and $H_3$
for hadrons and jets as function of energy or resolution, respectively.
One observes the approximate coincidence of the zeros of $H_2$ and the minima
near zero of $H_3$ corresponding to the ``Poissonian point''  and the 
alternating 
signs below and same sign above this point. Again, the agreement of the 
structures in the region of highly resolved jets is well reproduced and the 
same is observed for all available higher moments \cite{bfo}. As function of order $q$ 
one obtains for both hadrons and jets 
an oscillation pattern which depends on energy and resolution. The main 
characteristics can be derived from the numerical solution of the 
MLLA evolution equation and in good overall quantitative agreement with the 
ARIADNE-D MC. 

\section{Conclusions}
The perturbative approach to multiparticle production 
based on the MLLA is found to work well, 
also for the 
correlation phenomena discussed here, 
both for hadrons and jets at variable resolution
and it properly distinguishes between quark and gluon jets.
An important condition for this success is  the high 
accuracy of the calculation.
The DLA at realistic energies does not always give qualitatively
correct results and misses, for example, 
the Poissonian point and the oscillations of 
$H_q$ at higher energies. Also some additional terms in a $\sqrt{\alpha_s}$
expansion are insufficient for a quantitative description
of $N_g/N_q$ and the higher multiplicity moments. 
On the other hand, numerical solutions of the 
MLLA equation and the parton MC considered 
here come to a rather close description 
in terms of only two essential parameters $\Lambda,Q_0$.

The normalization parameter is found close to $K=1$ \cite{lo} 
which means that the 
global hadronic observables considered here 
can be described after replacing a hadron 
by a parton at resolution $Q_0\gsim\Lambda$. 
This description does not take into accont local effects like resonance 
production, but its success 
is in support of a 
 dual description of a large class of hadronic and partonic observables. 

\section*{Acknowledgements} 
I would like to thank Matt Buican, Clemens F\"orster
and Sergio Lupia for the enjoyable collaboration 
and to Lev Lipatov and Victor Kim for organizing this well inspired meeting 
and their kind hospitality during the visit.

\end{document}